\documentclass{article}
\usepackage[utf8]{inputenc}
\usepackage{soul}
\usepackage{xcolor}
\usepackage{url}
\usepackage{listings}
\usepackage[ruled,vlined]{algorithm2e}
\usepackage[colorinlistoftodos]{todonotes}
\usepackage{graphicx}
\usepackage{tabularx}
\usepackage{rotating}
\usepackage{tikz}
\usepackage{hyperref}
\usepackage{comment}
\usepackage{semantic}
\usepackage{subfigure}
\usepackage{multicol}
\usepackage{float}
\graphicspath{ {./FIG/} }
\title{Ontology-based Attack Graph Enrichment} 
\author{K. Saint-Hilaire$^1$, F. Cuppens$^2$, N. Cuppens$^2$, J. Garcia-Alfaro$^1$\\~~\\
\multicolumn{1}{p{.9\textwidth}}{\centering{$^1$Institut Polytechnique de Paris, T\'el\'ecom SudParis, France\\
$^2$Polytechnique Montr\'eal, Canada}}
}

\date{~~}

\newtheorem{mydef}{Definition}

\begin{document}

\maketitle
\noindent \textbf{Abstract}: Attack graphs provide a representation of possible actions that adversaries can perpetrate to attack a system. They are used by cybersecurity experts to make decisions, e.g., to decide
remediation and recovery plans. Different approaches can be used to build such graphs. We focus on logical attack graphs, based on predicate logic, to define the causality of adversarial actions. Since networks and vulnerabilities are constantly changing (e.g., new applications get installed on system devices, updated services get publicly exposed, etc.), we propose to enrich the attack graph generation approach with a semantic augmentation post-processing of the predicates. Graphs are now mapped to monitoring alerts confirming successful attack actions and updated according to network and vulnerability changes. As a result, predicates get periodically updated, based on attack evidences and ontology enrichment. This allows to verify whether changes lead the attacker to the initial goals or to cause further damage to the system not anticipated in the initial graphs. We illustrate the approach under the specific domain of cyber-physical security affecting smart cities. We  validate the approach using existing tools and ontologies.\\

\section{Introduction}
\label{sec:introduction}

Cybersecurity consists of protecting systems, networks, and programs against cyber-attacks that aim to access, modify or destroy sensitive information, extort money from users, or disrupt normal business processes. Cyber-attacks against networks are rising. Computer and network attacks and their countermeasures become more and more complicated~\cite{Roschke2009}. The understanding of attack realization against a system is essential. Attack graphs show possible paths that adversaries can use to reach their goals successfully.

There exist various types of attack graph models, mainly~\cite{Aguessy2017}: logical, topological, and probabilistic models. Logical models represent an attack as a logical predicate requiring successful preconditions for the attack to be perpetrated. This type of model accurately represents the process by which humans judge whether an attack is either possible or not. Topological models offer a higher-level view of possible attacks in an information system, representing an attack as a way of accessing new resources. Finally, probabilistic models, e.g., using Bayesian theories, assign probabilities to nodes and attack steps. 

In this paper, we choose to favor logical attack graphs. The rationale for our choice is as follows.
Both topological and probabilistic models provide less precision than logical models, f.i., in terms of explainability about how attacks were performed. Indeed, logical attack graphs illustrate the causes of the attacks instead of snapshots of the attack steps~\cite{Ou2006}. This offers several advantages. For instance, the size of the graph increases in a polynomial manner, whereas in other approaches it can increase exponentially. Moreover, in a logical attack graph, causality relations between adversaries and systems are already represented in the logical statements of nodes and edges. In the other approaches, one may need to go through Boolean variables to identify the cause of an adverse situation that allows adversaries' actions in a stage, hence increasing processing and inference complexity. In the case of logical attack graphs, the exploitation of existing vulnerabilities on an asset is the main cause of the attack.

Our work aims to tackle the following question: \emph{how can real-time system monitoring enrich a priori logical attack graphs by taking into account vulnerability and network configuration updates?} We claim that a posteriori  enrichment of the graphs would make possible to fulfill certain preconditions that were not taken into account in the generation of the initial graph. The new process may also allow to discover if the system is now exposed to different situations that can augment the attacks from the initial goals to other detrimental events, causing even further damages. The use of semantic information about system vulnerabilities leads our analysis.

Cybersecurity operators often rely on CVE (Common Vulnerability and Exposure) reports for information about vulnerable systems and libraries to prevent vulnerabilities exploitation. These reports
include descriptions, disclosure sources, and manually-populated
vulnerability characteristics. Characterizing software vulnerabilities is essential to identify the root cause of the vulnerability, as well as to understand
its consequences and attack mechanisms. The use of ontologies~\cite{uschold1996ontologies} is a proper way to
represent and communicate facts and relations between multiple
agents. Several ontologies describing collections of publicly known software vulnerabilities exist. Examples include the SEPSES (Semantic Processing of Security Event Streams) ontology~\cite{kiesling2019sepses}, which describes vulnerabilities extracted from the CVSS\footnote{\url{https://www.first.org/cvss/}} (Common Vulnerability Scoring System) database; and the Vulnerability
Description Ontology (VDO)~\cite{booth2016vulnerability}, proposed by the National Institute of Standards and Technology (NIST), in an effort to characterizing software vulnerabilities in a standard manner. The inference abilities of those existing ontologies justify their use for the enrichment of logical attack graphs. Moreover, it can help to guarantee that the attack graph remains faithful to system updates. This includes processing evidences of vulnerability exploitation, i.e., mapping of monitoring alerts against semantic ontologies.

To validate our approach, we conduct experimental work using the following setup. We use a scanner of vulnerabilities (based on Nessus Essentials\footnote{\url{https://www.tenable.com/products/nessus/nessus-essentials}}) to discover and list vulnerabilities in a given monitored system. The results are consumed by Mulval~\cite{Ou2005MulVALAL}, a logic-based attack graph engine.  We add system monitoring, using Prelude-OSS\footnote{\url{https://www.prelude-siem.com/en/oss-version/}}, an opens-source Security Information and Event Management (SIEM) system, enhanced with additional tools to trigger and post-process attack alerts. We also instantiate precise attacks to change the state of the system (i.e., exploitation of vulnerabilities) and use a recent implementation of VDO\footnote{\url{https://github.com/usnistgov/vulntology}} to enrich the initial attack graph by augmenting the predicates of the initial graph with the semantic data of VDO and the alerts from Prelude-OSS. Alerts trigger a search  within the graph, and expand those paths related to a successful vulnerability exploitation.

\medskip

\noindent \textbf{Paper Organization ---} Section~\ref{sec:background} provides a background of the subject and some preliminaries on the use of attack graphs. Section~\ref{sec:approach} presents our attack-graph enrichment approach. Section~\ref{sec:experiments} provides the experimental results. 
Section~\ref{sec:relatedwork} surveys related work. 
Section~\ref{sec:conclusion} concludes the paper.

\section{Background}
\label{sec:background}

\subsection{Literature on Attack Graphs}
Cyber-attacks are frequently represented in the information security literature as attack graphs. The idea is to identify all those potential paths that an adversary can take in order to perpetuate the exploitation of a series of vulnerabilities and compromise an information system. Different approaches exist, with respect to the way how we can construct and use such attack graphs.

Early literature on attack graphs used them to determine whether specific goal states can be reached by an adversary who is attempting to penetrate a system~\cite{Lippmann2005AnAR}. The starting vertex of the graph represents the initial state of the adversary in the network. The remaining vertices and edges may represent the actions conducted by the adversary, and the system changes due to the adversary actions. Actions may represent adversarial execution of vulnerabilities in the system. A series of actions may represent the adversary steps toward an escalation of privileges in the system, e.g., to obtain enough privileges on different devices or network components in the system. Actions can be combined using either OR (disjunctive) or AND (conjunctive) logic predicates~\cite{Schneier99}, as well as other attributes, such as the costs associated to the actions, their likelihood and probability of success, etc. In the end, a complete attack graph is expected to show all the possible sequences of actions that will allow the adversary to successfully perpetrate the attack (e.g., to penetrate into the system). Some other representations and uses are possible. For instance, instead of using vertices to represent system states and edges to represent attack actions, we can represent actions as vertices and system states as edges; instead of using single adversary starting locations, we can also assume multiple adversary starting locations or multiple targets and goals, etc.

Other models use directed graphs. For instance, topological attack graphs directly use topological nodes to represent information about systems' assets. The edges represent an adversary's steps to move from a topological parent node to a topological child node. The type of attack (exploitation of a vulnerability, credential thief, etc.) related to an attack step describes how the adversary can move between nodes. The set of conditions associated with an attack step depends on the type of attack.  A sensor can be associated with an attack step, a sensor that may alert that this attack has been detected. Similarly, probabilistic models using Bayesian networks can also represent attack graphs via directed acyclic graphs. Nodes represent random variables and edges represent probabilistic dependencies between variables~\cite{Aguessy2016}. An 
example is BAM (Bayesian Attack Model)~\cite{aguessy:hal-01144971}, which builds upon Bayesian attack trees. Nodes represent transitions, conditions, and sensors. Each node represents a Boolean random variable with two mutually exclusive states. A Bayesian transition node represents the random variable that describes the success or fail of a transition. A Bayesian condition node represents the random variable that describes if the condition is fulfilled. A Bayesian sensor node is a random variable that describes the state of a sensor. These nodes are linked with edges, which indicates that the child node conditionally depends on the state of its parents.

Compared to Bayesian networks, logical attack graphs provide some practical advantages. First, the use of acyclic graphs in Bayesian networks requires from heuristics to suppress paths that an adversary can follow. The inference of a Bayesian attack graph is very complex, since it needs to delve into the Boolean variables and follow several steps upstream to identify the adverse situation causes that enable an adversary's action at a stage. This leads to performance problems. In logical attack graphs, the causality is specified as graph edges. Thus, the inference of a logical attack graph is straightforward. Logical attack graphs are also more elaborated than topological attack graphs. 
In the sequel, we elaborate further on logical attack graph modeling.

\subsection{Logical Attack Graph Modelling}
\label{sec:preliminaries}

We define next some preliminary concepts such as Graph, Directed Graph, and AND-OR Graph, as underlying requirements for logical attack graph modeling~\cite{1386,Aguessy2017}.

\newtheorem{definition}{Definition}[section]

\begin{mydef}[Graph]\label{def:graph}
A Graph is a set $V$ of vertices, and a set $E$ of unordered and ordered pairs of vertices, denoted by $G(V;E)$. An unordered pair of vertices is an edge, while an ordered pair is an arc. A graph containing edges alone is non-oriented or undirected; a graph containing arcs alone is called oriented or directed.
\end{mydef}

\begin{mydef}[Directed Graph]\label{def:directedgraph}
A directed graph $G(V;A)$ consists of a non\-empty set $V$ of vertices and a set $E$ of arcs formed by pairs of elements of $V$. In a directed graph:
\begin{itemize}
    \item The parent or source of an arc $(v_1; v_2) \in A; v_1 \in V; v_2 \in V ,$ is $v_1$.
    \item The child or destination of an arc $(v_1; v_2) \in A; v_1 \in V; v2 \in V ,$ is $v_2$.
    \item The incoming arcs of a node $v$ are all the arcs for which v is the child: $\forall a = (v_1; v) \in A, with$ $v_1 \in V$.
    \item The outgoing arcs of a node $v$ are all the arcs for which v is the parent: $\forall a = (v; v2) \in A, with$ $v_2 \in V$.
    \item the indegree $deg^{\_}(v)$ of a vertex $v \in V$ is the number of arcs in A whose destination is the vertex $v$: $deg^{\_}(v)$= Card(\{$v_i; \forall v_i \in V; (vi; v) \in A$\}).
    \item the outdegree $deg^{+}(v)$ of a vertex $v \in V$ is the number of arcs in A whose destination is the vertex $v$: $deg^{+}(v)$= Card(\{$v_i; \forall v_i \in V; (vi; v) \in A$\}).
    \item a root is a vertex $v \in V$ for which $deg^{\_}(v)$ = 0 (no incoming arc).
    \item a sink is a vertex $v \in V$ for which $deg^{+}(v)$ = 0 (no outgoing arc).
\end{itemize}

\end{mydef}

\begin{mydef}[AND-OR Graph]\label{def:and-or-graph}
An AND-OR graph is a directed graph where each vertex $v$ is either an OR or an AND. A vertex represents a sub-objective and according to its type (AND or OR), it requires either the conjunction or disjunction of its children, to be fulfilled. A root node $n$ of an AND-OR graph can be called a precondition as it does not require any other node $n$ to be fulfilled.
\end{mydef}

According to Definitions~\ref{def:graph}, \ref{def:directedgraph}, and~\ref{def:and-or-graph}, logical attack graphs are based on AND-OR logical directed graphs. The nodes are logical facts describing adversaries' actions or the pre-requisites to carry them out. The edges correspond to the dependency relations between the nodes. Various operators can be taking in account in a logical attack graph depending on the approach. The more popular operators are AND and OR. AND operator describes the achievement's requirement of all the facts of its children for the logical fact of a node to be achieved. OR operator describes the achievement's requirement of at least one fact of its children for the logical fact of a node to be achieved.

\section{Our Approach}
\label{sec:approach}

We assume that after the generation of a (proactive) attack graph, using a priori knowledge about vulnerabilities and network data, both networks and vulnerabilities may evolve (i.e., the configuration of system devices may change, software updates may be enforces, etc.). Hence, the network is not exposed to the same vulnerabilities as the beginning of the attack graph generation process. It is essential to update the attack graph according to systems' changes. When updating a logical attack graph, causality relations between adversaries and systems shall be represented in the logical statements of nodes and edges. We propose a logical attack graph enrichment approach based on ontologies. Before moving forward with our approach, we start by introducing a representative use case that will help up to explain the rationale of our approach. Examples based on the use case scenario are used to exemplify how our approach conducts the generation and enrichment of attack graphs, as well as other tasks, such as periodic monitoring and ontology analysis.

\subsection{Use Case Scenario}
\label{sec:use-case}

This section describes a use case scenario provided by smart city stakeholders. We provide first the general context associated to the scenario, then we focus more in detail on possible attack consequences described by the stakeholders.

\subsubsection{General Description}

An infectious disease spreads across multiple continents. Health authorities  impose unexpected lockdowns on several countries. When the situation seems over, politicians decide to apply some unpopular restrictions, to prevent new spreading waves of the disease. The population gets furious. Violent groups connected through the internet take it as an opportunity to launch attacks against assets associated to public services. Their goal is to cause panic among the population. 

\subsubsection{Panic and Violence on a Transportation Service }

Politicians decide to engage in another period of lockdown. Protesters are loudly shouting outside a municipal building. Social media respond positively to the movement. A mass of citizens arrives at the location. Public transportation in the area is heavily affected, causing long delays. Tensions and altercations rise with the increase of protesters. A fake alert, pretending to come from the municipality network, force people to evacuate the area. People get injured. Images and videos of altercations, evacuation, and car fires are posted on social media. At the same time, a denial-of-service cyber-attack against the municipality network is perpetrated. Machines and sensors get out-of-service, causing further delays in the transportation service of the city. People trying to leave the area start fighting, forcing the authorities to close all transportation services. The mass of people in a given bus affects the health of several passengers.

\subsection{Generation of the Attack Graph}

The generation of a logical attack graph requires the definition of rules describing causality relations. As an example, we consider code execution. Code execution on a machine allows an adversary to have access to a host. This scenario corresponds to the logical implication detailed by the following rule:
\begin{center}
\framebox[11cm][l]{
\begin{footnotesize}
\begin{minipage}{11cm}
$execCode(h,a) \rightarrow canAccesHost(h)$
\end{minipage}
\end{footnotesize}
}
\end{center}
where $canAccesHost(h)$ is a logical rule describing the accessibility to host $h$, and $execCode(h,a)$ a fact assessing that an adversary $a$ executed code in $h$. The example can be extended as follows:
\begin{center}
\framebox[11cm][l]{
\begin{footnotesize}
\begin{minipage}{11cm}
$execCode(h,a) \land hasCredentialsOnMemory(h,u)  \rightarrow harvestCredentials(h,u)$
\end{minipage}
\end{footnotesize}
}
\end{center}
where $harvestCredentials(h,u)$ describes a series of credentials harvesting on host $h$, $execCode(h,a)$ the fact that an adversary $a$ is executing code on host $h$, and $hasCredentialsOnMemory(h,u)$ the fact of storing the credentials on the memory of host $h$. The example describes the fact of an adversary harvesting the credentials of a previous user that logged onto the system, by finding them in the memory of that precise system. 

\begin{comment}
\begin{algorithm}[H]
\caption{}
\label{alg:alg7}
$canAccesHost(h)$: Rule describing accessibility to host $h$.

$execCode(h,a)$: Fact assessing that an adversary $a$ executed code in $h$.

\[
   execCode(h,a) \rightarrow canAccesHost(h)
\]
\end{algorithm}
\end{comment}

\begin{comment}
\begin{algorithm}[H]
\caption{}
\label{alg:alg8}
$harvestCredentials(h,u)$: Rule describing credentials harvesting on host h.

$execCode(h,a)$: Fact that the adversary a is executing code on the host h.

$hasCredentialsOnMemory(h,u)$: The credentials are saved on the memory of host h.

\[
   execCode(h,a) \land hasCredentialsOnMemory(h,u)  \rightarrow harvestCredentials(h,u)
\]
\end{algorithm}
\end{comment}

\subsection{Monitoring the Information System}

In order to update the attack graph based on the real-time state of the system, we can also monitor the information system. The output of the monitoring process can get continuously mapped with the initial nodes of the attack graph, in order to find out if a vulnerability is being exploited. The mapping between the monitored system and the attack graph is described bellow:

\begin{center}
\framebox[11cm][l]{
\begin{footnotesize}
\begin{minipage}{11cm}
$\forall{n \in N}: (vulExists(h,x,y,z) \land networkServiceInfo(h,s,p,a,u) \rightarrow F_{1}$
\end{minipage}
\end{footnotesize}
}
\end{center}
where $(vulExists(h,x,y,z)$ describes the existence of a vulnerability $x$ on host $h$ which allows action $y$ resulting in $z$. Likewise,  $networkServiceInfo(h,s,p,a,u)$ describes that user $u$ has a session open on host $h$ where a given service product $p$ is installed, using port $a$ and protocol $p$.
 
The evaluation of a successful mapping implies finding further details about specific vulnerabilities. We propose to use a vulnerability ontology to conduct such a process, represented by $F_1$. Next, we provide some more details about this process using semantic information about concrete vulnerabilities.

\begin{comment}
\begin{algorithm}[H]
\caption{Mapping of the system and the attack graph}
\label{alg:alg0}
$N$: Nodes in attack graph.

$vulExists(h,x,y,z)$: Vulnerability $x$ exists on host $h$ and allows the exploit $y$ that can result in $z$.

$networkServiceInfo(h,s,p,a,u)$: The user $u$ has a session opened on host $h$ where product $p$ is installed and port $a$ is opened for protocol $p$.

$F_1$: Look for the vulnerability characteristics in the vulnerability ontology

\[
   \forall{n \in N}: (vulExists(h,x,y,z) \land networkServiceInfo(h,s,p,a,u)) \rightarrow F_{1}
\]
\end{algorithm}
\end{comment}

\subsection{Vulnerabilities and Ontologies}

Vulnerability information is necessary for both the attack graph generation process and the updates. Vulnerabilities enable the adversary to take actions towards the initial adversarial goals, or alternative actions affecting the system in different ways. Lists of uniquely identifiers in CVE (Common Vulnerabilities and Exposures), a collection of publicly known software vulnerabilities, are complemented with valuable descriptions about the vulnerability, its preconditions and post-conditions, and practical ways to be exploited. The information contained in CVE's descriptions can also lead to other valuable characterizations, for example, impact to the system if the vulnerability is exploited.

An ontology is a formal description of a field of knowledge and is represented by descriptive logic. An ontology brings semantic support and unifies unstructured data. Ontologies have been widely used in the field of cybersecurity, for instance, to represent vulnerability classes and their inner relations. Table~\ref{table:CVE-2002-0392} shows an example, representing the classification of a given vulnerability listed in CVE (with identifier CVE-2002-0392). Ontologies also offer inference abilities, which we will use to enrich logical attack graphs when the exploitation of a vulnerability is being reporting during the monitoring process of a vulnerable system.

\begin{table}[h!]
\fontsize{7.5}{10}\selectfont
\begin{tabularx}{1\textwidth} { 
  | >{\raggedright\arraybackslash}X 
  | >{\raggedright\arraybackslash}X 
  | >{\raggedright\arraybackslash}X
  | >{\raggedright\arraybackslash}X
  | >{\raggedright\arraybackslash}X | }
 \hline
 CVE-ID & Product & Type & Action & Impact\\
 \hline
 CVE-2002-0392 & Apache  & remote  & Code Execution & Privilege Escalation\\
\hline
\end{tabularx}
\caption{Classification of CVE-2002-0392 characteristics.}
\label{table:CVE-2002-0392}
\end{table}

\begin{algorithm}[b]
\caption{Enrichment of a proactive attack graph based on a vulnerability ontology and monitored system information}
\label{alg:alg1}
$h_1$: A threat exists on a vulnerable component of the monitored system.

$h_2$: Post-conditions of the exploited vulnerability are found in the ontology.

$P_1$: Add new path on the attack graph.
\[
   (h_{1}\land h_{2}) \rightarrow P_{1}
\]
\end{algorithm}

\begin{algorithm}[t]
\caption{Inference rule for mass on buses scenario}
\label{alg:alg2}
$v_1$: Node corresponds to reboot of a machine.

$v_2$: Node corresponds to mass on buses.

The child or destination of an arc $(v_1; v_2) \in A; v_1 \in V; v_2 \in V ,$ is $v_2$.

\[
   (v_{1} \land v_{2}) \rightarrow (v_{1}; v_{2})
\]

The inference rule is:
\[
\inference {v_{1} \quad v_{2}}{(v_{1}; v_{2})}[r]
\]
\end{algorithm}

\subsection{Enrichment of Attack Graphs}

Algorithm~\ref{alg:alg1} represents our proposed approach for enriching attack graphs based on a vulnerability ontology and monitoring system information. When a threat exists on a vulnerable component of the monitored system, it is necessary to look through the vulnerability characteristics to find its post-conditions. Those post-conditions enrich the attack graph with new paths. The inference rules allow knowing what those new paths can bring to the attack goal. As an example,  Algorithm~\ref{alg:alg2} shows an inference rule based on the definition of a Directed Graph in Definition~\ref{def:directedgraph}, and deduces the consequence of restarting a device in the scenario of Section~\ref{sec:use-case} (i.e., a cyber-attack on a municipality network that takes a given device out-of-service for a while). During the attack, the lack of communication between an application and a server causes a problem in the logistics of the transportation service. There is a delay in the bus service. This scenario is not anticipated in the initial graph. It is necessary to update the graph and add the new path that allows the adversary to reach the goal (i.e., cause panic and violence of people). Updating the graph will help the operators to inform the authorities and explore the best remediation strategy to mitigate the damages as soon as possible. Therefore, it is necessary to define inference rules like the one shown in Algorithm~\ref{alg:alg2}, to update the graph in such a situation.

\section{Implementation}
\label{sec:experiments}

\subsection{Setup}
\label{sec:setup}

\begin{figure}[!t]
\centering
\includegraphics[scale=0.34]{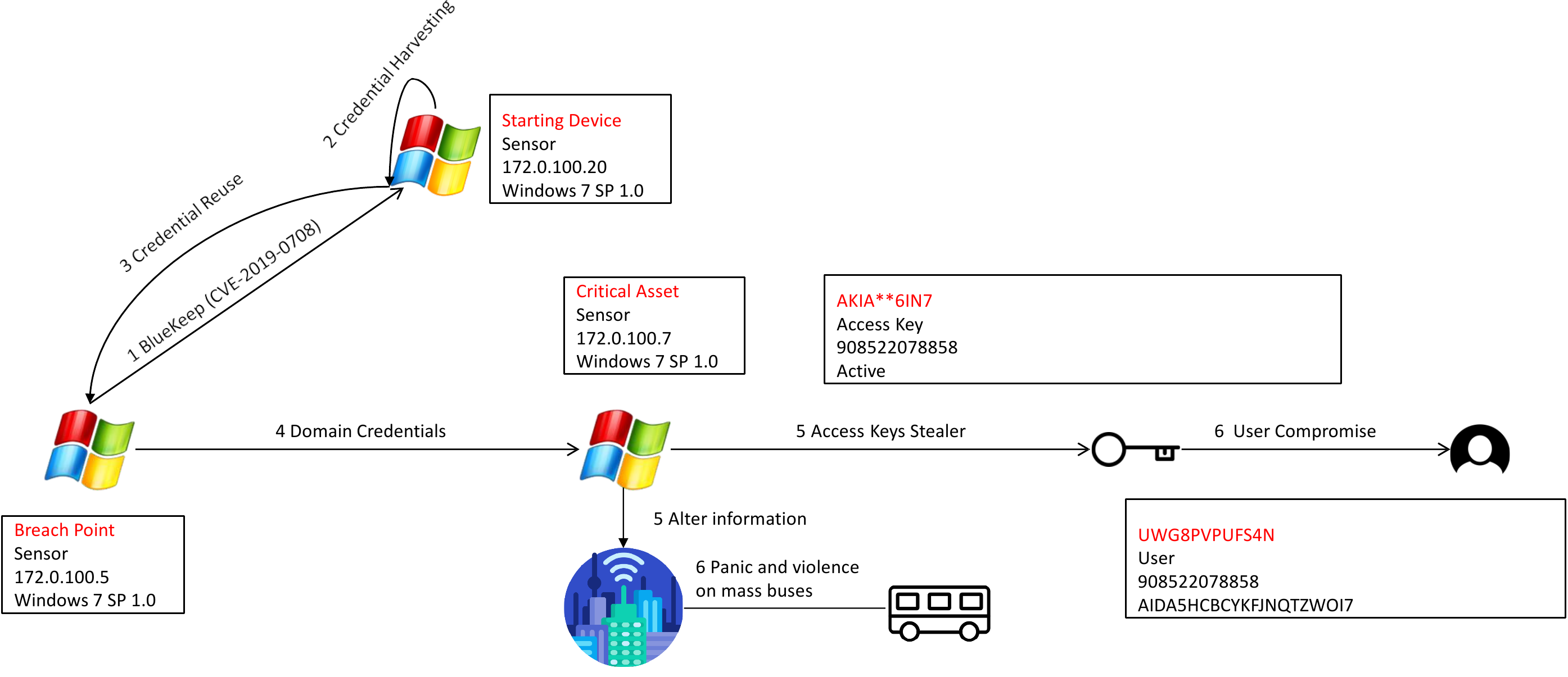}
\caption{Cyber-physical attack scenario. An adversary exploits the vulnerability associated to CVE-2019-0708 on a Starting Device. Then, administrator credentials are harvested from the memory of the device, and reused by the adversary to take control over a critical asset. The attack affects both physical and digital elements associated to the system (e.g., people and services).}
\label{fig:cps}
\end{figure}

In order to validate our approach, we instantiate the scenario depicted in Figure~\ref{fig:cps}. It represents a cyber-physical system monitored by a Security Information and  Event Management (SIEM) system, based on Prelude-OSS\footnote{\url{https://www.prelude-siem.com/en/oss-version/}}. We use a virtual machine representing the starting device of the scenario, another machine to instantiate the breach point, and a third one representing the critical asset. We use Nessus Essentials\footnote{\url{https://www.tenable.com/products/nessus/nessus-essentials}} to discover and list vulnerabilities in the monitored system. Data from  Nessus is consumed by MulVAL~\cite{Ou2005MulVALAL}, a reasoning engine based on logical programming, to generate a logic-based attack graph. We also use a practical implementation\footnote{\url{https://github.com/usnistgov/vulntology}} of NIST's Vulnerability
Description Ontology (VDO)~\cite{booth2016vulnerability}, and Prelude-OSS to map the information contained in VDO into the attack graph, upon reception of Prelude-OSS' alerts. 

The rationale of the scenario depicted in Figure~\ref{fig:cps} is as follows. An adversary succeeds to execute arbitrary code on the starting device by connecting remotely through RDP (Remote Desktop Protocol, a network service that provides users with graphical means to remotely control computers). The adversary can then read the memory of the starting device. The credentials of the administrator are saved in the memory of the starting device. Then, the adversary harvests those credentials. We assume that the administrator can connect to all the machines in the domain, to remotely manage them. Then, an adversary capable of reusing the credentials can log onto the breach point and remotely connect to the critical asset. The adversary also perpetrates a DNS Poisoning attack~\cite{hu2018measuring}, in order to eavesdrop network traffic. The adversary also perpetrates some integrity attacks, in order to modify application level information, such as the bus schedule and routes, to perturb the influence of traffic and cause a congestion increase. This causes citizens taking the wrong buses at the wrong time, leading into the situation of panic and violence mentioned in Section~\ref{sec:use-case}. In parallel, the adversary reuses the domain credentials to steal some other access keys and impersonate other users (shown in Figure~\ref{fig:cps} with steps \emph{Access Keys Stealer} and \emph{User Compromise}). This parallel scenario leads to the exploitation of other vulnerabilities and an eventual denial-of-service attack.

\subsubsection{MulVAL}
\label{sec:mulval}
Based on the scenario shown in Figure~\ref{fig:cps}, we create input data for MulVAL, as well as interaction rulesets associated to VDO. We encode the new interaction rules as Horn clauses~\cite{Ou2005MulVALAL}. The first line corresponds to a first-order logic conclusion. The remaining lines represent the enabling conditions. The clauses below corresponds to the following statement from the scenario shown in Figure~\ref{fig:cps}: \emph{'the breach point credentials can be harvested on the starting device only if there is previously an execution code exploit on the starting device and the credentials of the administrator are saved onto the memory of the starting device'}.

\begin{verbatim}
harvestCredentials(_host, _lastuser) :-
       execCode(_host, _user),
       hasCredentialsOnMemory(_host, _lastuser)
\end{verbatim}

The clauses below represent the following facts: \emph{'it is possible to log into the breach point with the administrator credentials when these credentials have been harvested and because the breach point and the starting device are on the same network and can communicate through a given protocol and port}.

\begin{verbatim}
logOn(_host, _user) :-
        networkServiceInfo(_host, _program, _protocol, _port, _user),
        hacl(_host, _h, _protocol, _port),
        harvestCredentials(_h, _user)
\end{verbatim}

\subsubsection{Ontology}
\label{sec:ontology}
\begin{figure}[h]
\centering
\includegraphics[scale=0.5]{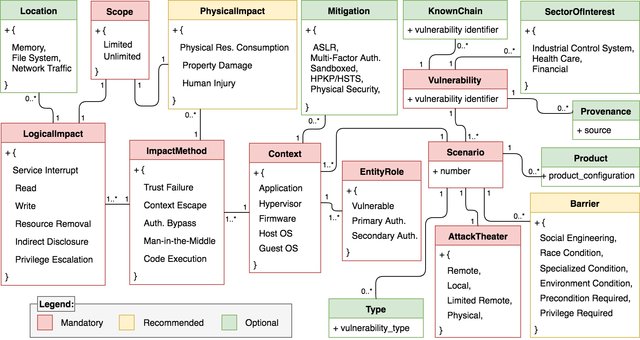}
\caption{The NIST Vulnerability Description Ontology (VDO)~\cite{Gonzalez}. This figure represents the different classes of NIST vulnerability ontology with their properties.\label{fig_vdo}}
\end{figure}

We use VDO, an ontology of CVEs proposed by NIST.  Figure~\ref{fig_vdo}, from~\cite{Gonzalez}, represents various attributes of the VDO ontology, for characterization of software vulnerabilities. Various attributes, such as \textit{Attack Threater}, \textit{Impact Method} and \textit{Logical Impact} are mandatory. The value of \textit{Attack Threater} characterizes the area or place from which an attack must occur. \textit{Impact Method} describes how a vulnerability can be exploited. \textit{Logical Impact} describes the possible impacts a successful exploitation of the Vulnerability can have. For each CVE affecting the monitored system, we can fulfill the classes of information from the ontology, according to the description and metrics of the CVE.\\

A simplified description of \emph{CVE-2019-0708} according to VDO would state, among other details, that '\emph{a remote code execution vulnerability exists in Remote Desktop Services, formerly known as Terminal Services, which can be exploited by an unauthenticated attacker connecting to the target system using TCP or UDP traffic and sending specially crafted requests}.

\begin{table}[h!]
\centering
\includegraphics[scale=0.72]{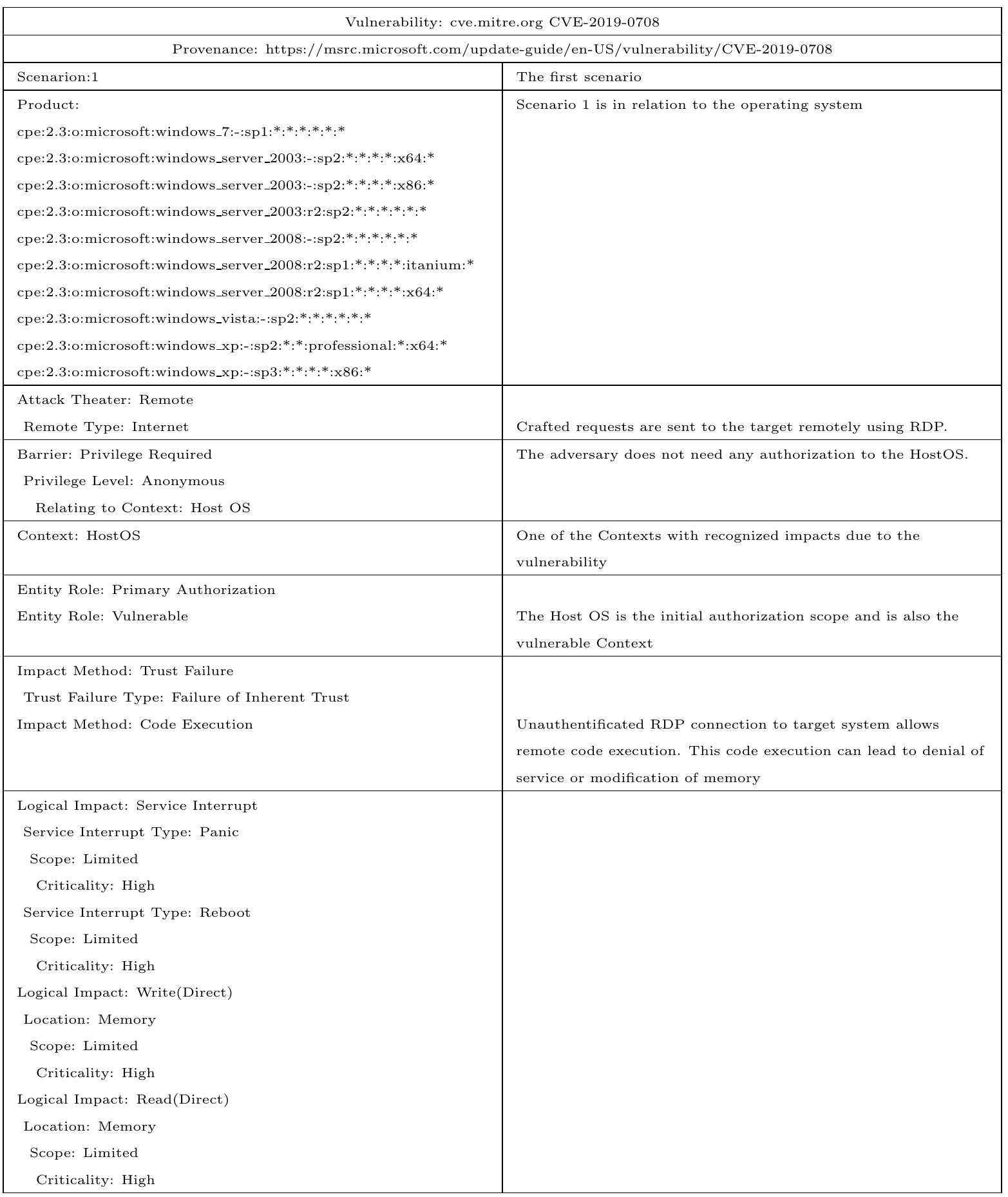}
\caption{Attributes associated with CVE-2019-0708.}
\label{table:1}
\end{table}

\subsubsection{Prelude-ELK}
\label{sec:preludeelk}

We use an extended version of Prelude-OSS with ELK (Elasticsearch, Logstash, and Kibana). The code is available online\footnote{\url{https://github.com/Kekere/prelude-elk}}. Elasticsearch allows us indexing and processing unstructured data. It also provides a distributed web interface to access the resulting information. Logstash is the parsing engine associated with Elasticsearch for collecting, analyzing, and storing logs. It can integrate many sources simultaneously. Finally, Kibana is a data visualization platform that provides visualization functionalities on indexed content in Elasticsearch. Users can create dashboards with charts and maps of large volumes of data. 

The addition of the ELK stack into Prelude-OSS allows the injection and visualization of third-party logs received from both system and network components via TCP/IP messages. The collection of data can still be combined with the classic collection and visualization tools of Prelude-OSS. For instance, we can keep using Prelude’s LML (Log Monitoring Lackey) and third-party sensors to monitor and process Syslog messages generated from different hosts on heterogeneous platforms. In addition, we also extend Prelude-OSS with Suricata\footnote{\url{https://suricata.io/}}, as a third-party sensor reporting alerts on the exploitation of known vulnerabilities. We install Rsyslog Windows Agent~\footnote{https://www.rsyslog.com/windows-agent/} and Suricata on each of the virtual machines, in order to monitor them with the ELK extension of Prelude-OSS. Logstash inserts the alerts into Elasticsearch (in JSON format). The results are processed in real-time, mapping the alerts and VDO's data while conducting our attack graph enrichment process.

\subsubsection{Web Interface}
\label{sec:webinterface}
We create a web interface using PHP, Javascript, JQuery, D3.JS, and HTML, where we upload the XML outputs of Mulval. The inference engine converts the XML into JSON. From this JSON, the engine displays a web visualization of the attack graph. The server consults the Elasticsearch index in real-time. The last alert's IP address, port, and protocol match the attack graph. When a vulnerability is likely to be exploited, the engine consults the vulnerability ontology to find other post-conditions for the vulnerability. The tool updates the attack graph according to the ontology.

\subsection{Results}
\label{sec:results}

\begin{figure}[hptb]
\centering
\subfigure[Initial attack graph. The adversary gains network access on Node $25$. When all the preconditions are met on each stage, the adversary can take actions represented by green nodes to reach Node $1$, which represents the adversarial goal (i.e., panic and violence on mass buses).\label{fig:before}]{
\includegraphics[width=\textwidth]{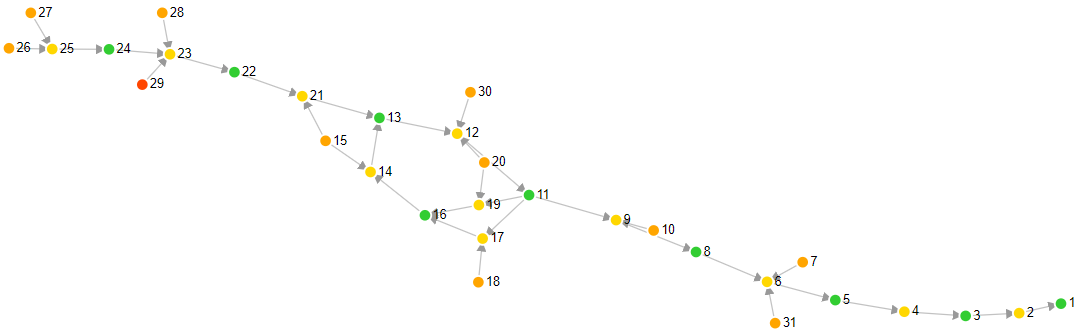}
}
\subfigure[Enriched attack graph. A new path towards Node $1$ (panic and violence on mass buses), is discovered using the ontology. The adversary can now take a much shorter path to reach the final goal. \label{fig:after}]{
\includegraphics[width=0.6\textwidth]{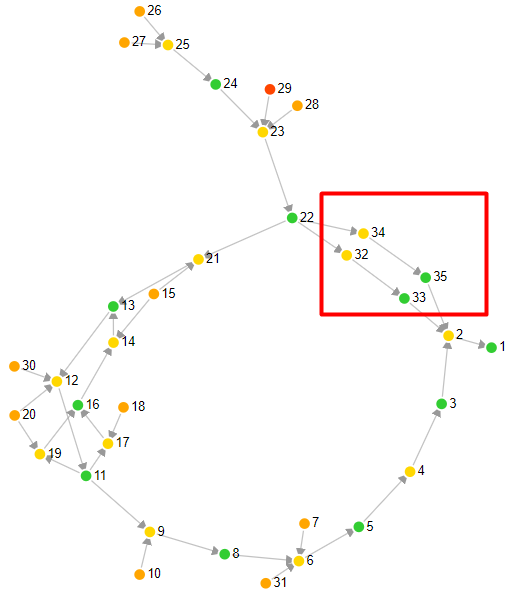}
}
\caption{Sample results. (a) Attack graph generated for the scenario of mass on buses. (b) The same attack graph, once enriched with data from the ontology. In both graphs, a red node represents the existence of a vulnerability on a given resource. An orange node represents network configuration (e.g.,  characteristic of a machine or connection between two machines in the network). When preconditions are respected, a yellow node represents an inference rule that leads to a fact (represented by a green node). \label{fig:before-after}}
\end{figure}

Figure~\ref{fig:before}, represents the attack graph generated for the scenario depicted in Figure~\ref{fig:cps}. The goal, represented by Node $1$, is to cause panic and violence (see use-case scenario described in Section~\ref{sec:use-case}). A red node represents the existence of a vulnerability on a device. An orange node represents network configuration, e.g., characteristics of a device, connection between two deices in the network, etc. When the preconditions are respected, a yellow node represents the inference rules leading to a fact. Facts are represented by green nodes. For instance, Node $26$ represents remote connectivity of the Starting Device in Figure~\ref{fig:cps}, which can be remotely accessed using RDP (Remote Desktop Protocol) services. Node $27$ concerns the location of the adversary in the network. Node $25$ represents the rule that leads the adversary to gain direct network access (i.e., Node $24$) on the starting device, when preconditions on Nodes $26$ and $27$ are met. Node $29$ concerns the existence of the vulnerability \emph{CVE-2019-0708} on the starting device. \emph{CVE-2019-0708} consists of a remote code execution vulnerability. Node $28$ concerns the network configuration of the Starting Device. Some other practical details encoded in the graph correspond to the operating systems (Windows 7), open TCP ports (3389), and identity of the user at the Starting Device (username \emph{olivia}). Finally, Node $23$ has Nodes $24$, $28$, and $29$ as main preconditions.

Alerts are processed with Prelude-ELK (cf. Section~\ref{sec:preludeelk}) in real-time. The inference engine finds exploited nodes based on network information associated to a victim device (i.e., the Starting Device in  Figure~\ref{fig:cps}), such as IP address, protocol, and port.  The service consults VDO (i.e., our vulnerability ontology) to find other post-conditions associated with \emph{CVE-2019-0708}. For instance, it compares the operating system associated to the victim device against the list of products listed in \emph{CVE-2019-0708}. As a result, an enriched attack graph is derived. Figure~\ref{fig:after} represents such an enriched attack graph. The four nodes highlighted with the red square correspond to the new nodes added to the enriched attack graph. They represent the logical impacts derived from the ontology. Node $33$ describes the Starting Device being restarted. Node $35$ describes a system crash of the Starting Device (i.e., Starting Device stops functioning properly). The consequence of Nodes $33$ and $35$ (i.e., Starting Device restarting or unavailable) leads to the mass on buses scenario (i.e., by inference, Nodes $33$ and $35$ are targeting now Node $2$, which is the rule concerning mass on buses). In Figure ~\ref{fig:after}, the four added nodes represent a new path that the adversary can take to cause panic and violence. As we can see, the enriched graph is now an acyclic graph. The new path is shorter than the predicted one. The adversary can reach the goal, represented by Node $1$, much sooner than expected. This difference would makes operators aware that it is more urgent to apply a remediation plan.

\section{Related Work}
\label{sec:relatedwork}

\subsection{Attack Graph Generation Approaches}

Work by Ghosh and Ghosh~\cite{Ghosh2012} propose a planning-based approach for attack graph generation and analysis. In this approach, initial network configuration and description of exploits serve as input for the minimal attack graph generation. Shirazi et al.~\cite{B2019} present an approach for modeling attack-graph generation and analysis problems as a planning problem. They present a tool called AGBuilder that generates attack graphs using the Planning Domain Definition Language (PDDL) from extracted vulnerability information. 

Roschke et al.~\cite{Roschke2009} present an approach of vulnerability information extraction for attack graph generation using MulVAL. The proposal integrated attack graph workflows with SIEM alerts in terms of data fusion and correlation. Alerts are filtered based on vulnerability and system information of the attack graph. The correlation process can reveal a new way the network can be attacked. In such a case, the attack graph is updated. 

Compared to those aforementioned approaches, we monitor the network to update the attack graph based on state change of the network and generate attack graphs based on network information received from Nessus scans. We also enrich the attack graph based on vulnerability information from CVEs and alerts received from a SIEM. We use a logical attack graph generation approach instead of a planning-based attack graph generation one. With a logical approach, the inference is more straightforward. Moreover, the semantics abilities enhance attack graph enrichment with ontology. We use a vulnerability ontology to correlate alerts with the system and vulnerability information.

\subsection{Ontology and Attack Graph Generation}

Falodiya et al.~\cite{Falodiya2018}  propose an ontology-based approach for attack graphs. The idea is to use an exploit dependency attack graph, an equivalence of logical attack graphs. The work presents an algorithmic solution to traverse the attack graph and add the extracted data into the ontology. 

Lee et al.~\cite{Lee2019} also propose an approach for converting an attack graph into an ontology. Their formalism is based on an attack-graph approach by Ingols et al.~\cite{Ingols}. The extraction of semantics from the graph is then labeled to build an RDF (Resource Description Framework) graph. Using RDF schema is beneficial for inferences from data and enhance searching. Wu et al.~\cite{Wu2018} propose as well an attack graph generation approach based on the inference ability of cybersecurity ontologies. 

In our approach, we use these abilities of semantic languages to enrich logical attack graphs easily. We use a NIST standardized ontology (VDO, for Vulnerability Description Ontology) as the primary source of such vulnerability semantics. VDO corresponds well with the logical attack graph approach. It provides mandatory classes such as \textit{Logical Impact} and \textit{Product}, which we use to map alerts with attack graph nodes. New attack paths can be discovered when looking for other logical impacts of a given CVE in VDO. With this approach, we avoid recomputing the attack graph from scratch in the reasoning engine each time the system state change. The semantic abilities of logical attack graphs and ontologies also allow us to take an incremental approach to update the graphs. This improves the automation of the enrichment process (i.e., cybersecurity operators do not have to manually modify inputs to update the attack graphs).

\section{Conclusion}
\label{sec:conclusion}
We have proposed an ontology-based approach for attack graph enrichment. We use logical graph modeling, in which attacks are represented with predicates. Successful precondition validation represents successful attack perpetration.  Compared to other similar approaches, such as topological and probabilistic attack graphs, our approach simplifies the inference process, since graphs' edges specify now causality. We have implemented the proposed approach using existing software. We have validated the approach based on a cyber-physical use-case, proposed by smart-city stakeholders. We have validated the full approach, from the generation of an initial attack graph (using network vulnerability scans), to the enrichment of the graph (mapping monitoring alerts and ontology semantics in real-time). The predictions of the initial graph get successfully updated into the enriched graph, based on attack evidences and semantic augmentation.\\

\noindent \textbf{Acknowledgements ---} We acknowledge financial support from the European Commission (H2020 IMPETUS project, under grant agreement 883286).

%\bibliographystyle{plain}
%\bibliography{biblio}

\end{document}